\documentclass[conference]{IEEEtran}

\usepackage{cite}
\usepackage{amsmath,amssymb,amsfonts}
\usepackage{algorithmic}
\usepackage{graphicx}
\usepackage{textcomp}
\usepackage{xcolor}
\usepackage{url}

\def\BibTeX{{\rm B\kern-.05em{\sc i\kern-.025em b}\kern-.08em
    T\kern-.1667em\lower.7ex\hbox{E}\kern-.125emX}}

\usepackage{fancyhdr}   %Kopfzeilen und co.
\fancypagestyle{firstpage}{     % define a custom header (Kopfzeile)
  \fancyhf{}
  \chead{\textcolor{gray}{This article has been accepted for publication in the flagship conference of \\ the IEEE International Instrumentation and Measurement Technology Conference(I2MTC).}}
  \fancyfoot[C]{\small{\textcolor{gray}{~\copyright~ 2024 IEEE.  Personal use of this material is permitted.  Permission from IEEE must be obtained for all other uses, in any current or future media, including reprinting/republishing this material for advertising or promotional purposes, creating new collective works, for resale or redistribution to servers or lists, or reuse of any copyrighted component of this work in other works.}}}

}

\begin{document}

\title{CycloWatt: An Affordable, TinyML-enhanced IoT Device Revolutionizing Cycling Power Metrics\\}

\author{\IEEEauthorblockN{1\textsuperscript{st} Victor Luder}
\IEEEauthorblockA{\textit{DITET} \\
\textit{ETH Zurich}\\
Zurich, Switzerland \\
luderv@ethz.ch}
\and
\IEEEauthorblockN{2\textsuperscript{nd} Sizhen Bian}
\IEEEauthorblockA{\textit{PBL-DITET} \\
\textit{ETH Zurich}\\
Zurich, Switzerland \\
sizhen.bian@pbl.ee.ethz.ch}
\and
\IEEEauthorblockN{3\textsuperscript{rd} Michele Magno}
\IEEEauthorblockA{\textit{PBL-DITET} \\
\textit{ETH Zurich}\\
Zurich, Switzerland \\
michele.magno@pbl.ee.ethz.ch}
}
\maketitle

\begin{abstract}
Cycling power measurement is an indispensable metric with profound implications for cyclists' performance and fitness levels. It empowers riders with real-time feedback, supports precise training regimen planning, mitigates injury risks, and enhances muscular development. Despite these advantages, the widespread adoption of cycling power meters has been hampered by their prohibitive cost and deployment complexity. This paper pioneers a groundbreaking approach to power measurement in cycling, prioritizing affordability and user-friendliness. To achieve this goal, we introduce a cutting-edge Internet of Things (IoT) device that seamlessly integrates force signals with inertial sensor data while leveraging the power of edge machine learning techniques. In-field experimental evaluations demonstrate that our prototype can estimate power with remarkable accuracy, boasting a Mean Absolute Error (MAE) of only 12.29 Watts (4.1\%). Notably, our design emphasizes energy efficiency, operating in a low-power mode that consumes a mere 50 milliwatts and offers an exceptional battery life of up to 25.8 hours in always-on active mode. With an ultra-low latency of 4.33 milliseconds for data processing and inference, our system ensures real-time power estimation during cycling activities. Incorporating IoT concepts and devices, this paper marks a significant milestone in developing cost-effective and accurate cycling power meters. 
\end{abstract}


\section{Introduction}

\thispagestyle{firstpage}

A cycling power meter provides users with immediate feedback on their fitness level and facilitates performance evaluation~\cite{broeder2013power}. With accurate real-time baseline data, cyclists can determine race strategy, pacing, and tactics during training and competitions. Besides that, by providing a foundation for precise training session planning, risks of injuries associated with excessive strain during workouts can be reduced~\cite{leo2022power}. The cycling power meter is thus a tool of utmost importance for professional and amateur cyclists~\cite{passfield2017knowledge}. 
However, existing power meters suffer from certain drawbacks. Chiefly, they are limited to a single bike due to a semi-permanent attachment~\cite{rodriguez2021assioma, granier2020validity}. This is because the applied torque must be measured, which mandates the integration of sensors into the bike. Thus, switching the sensing unit between bikes requires time, expertise, and specialized tools~\cite{AIEndurance}. Moreover, the high price point of these devices poses a considerable obstacle. Our questionnaire survey with 154 participants, achieving a confidence level of 95\% and a margin of error of 7.9\%, reveals that while 83\% of cyclists express interest in owning a power meter, 57\% are unwilling to pay the current market price. As a result, power meters remain inaccessible to a significant portion of amateur cyclists.

\begin{table*}[!t]
\centering
%\begin{threeparttable}
\caption{Overview power estimation in cycling}
\label{relatedwork}
\begin{tabular}{ p{1.3cm} p{3.7cm}  p{2.8cm} p{3.4cm} p{2.8cm} }
%\toprule
\hline
\textbf{Authors/ year} & \textbf{Application} &  \textbf{Data} & \textbf{Error} & \textbf{Computing model} \\
%\midrule
\hline
~\cite{accuratePowerMeterDesign}-2017 & Commercially available power meter & torque, IMU & 
2.5\% & Mathematical model\\
\hline
\cite{PowerwithML}-2020 & Average power estimation & heart-rate, cadence, ridden kilometers & 20 Watts Mean Absolute Error & Cluster based regression model\\
\hline
\cite{PresentationPowerEstimation}-2018  & Power estimation in cycling & heart-rate, speed, cadence, distance, elevation &  25.1 Watts Mean Absolute Error & Deep neural network \\
\hline
\cite{AIEndurance}-2023 & AI based training support &  heart-rate, gradient, cadence, aerobic threshold estimation  &  15\% - 25\% & - \\
\hline
\cite{MLinCycling}-2017 & Predict heart-rate response &  time, speed, distance, altitude, power, weight & -  & Recurrent neuronal network (RNN) \\
\hline
\textbf{Our solution} & \textbf{On the edge cycling power monitoring}   & \textbf{force, cadence, IMU}  & \textbf{15.32 Watts Mean Absolute Error (5.1\%)} & \textbf{Dense Neural network} \\
\hline
%\bottomrule
\hline
\end{tabular}
\end{table*}

Current industry methodologies primarily involve utilizing resistive strain gauges to measure the force in various components such as pedals, crank arms, crank spiders, chains, or rear wheel hub, and using simple statistical methods for power estimation~\cite{allen2019training, accuratePowerMeterDesign}. The industry standard attests to a precision of approximately 2.5\%~\cite{accuratePowerMeterDesign}. However, practical evaluations still show that extra calibrating and adjusting are needed for reliable measurement with different cycling environmental conditions~\cite{maier2017accuracy}.
On the other side, the research community has exhibited a growing interest in the past years, exploring the potential of power estimation in cycling using machine learning skills, aiming to unveil novel, reliable, and simplified designs ~\cite{PowerwithML,MLinCycling}. 

The rising prominence of Internet of Things (IoT) devices has fueled a remarkable surge in interest in local data processing through machine learning~\cite{hu2020fast, bonazzi2023low, zhou2023one, sz2021capacitive}. This trend has garnered even greater attention due to notable advancements in sensors, low-power electronics, and wireless technologies~\cite{giordano2022survey, bonazzi2023tinytracker}. Among the plethora of potential applications benefiting from machine learning on the edge, power estimation in cycling serves as an excellent example of a real-life implementation. By continuously capturing and analyzing data during cycling sessions, sensor data can be processed at its collection to provide instantaneous feedback on the exerted power. 
%However, this endeavor is not without challenges. Continuously capturing and analyzing data during cycling sessions presents hurdles in real-time data processing with low-power processors. The demand for processing sensor data at the point of collection to provide instantaneous feedback on the exerted power, necessitates efficient and low-latency machine learning models. In this context, advancements in sensor technology and algorithm optimization, such as quantized neural networks or embedded signal processing are vital to meet the stringent real-time requirements of power estimation, while also addressing the constraints of edge computing devices.

This paper represents a case study of lightweight, energy-efficient, and low-latency Tiny Machine Learning (TinyML) models tailored for edge processors, specifically applied to the domain of power estimation in cycling. In particular, we present the conceptualization, development, and real-world implementation of CycloWatt, an IoT device meticulously engineered to be energy-efficient and long-lasting.

%Our work marks a typical use case of AI-IoT, spearheaded by its novel approach to enhancing cycling power metrics. This endeavor is fueled by an ardent commitment to surmounting the design and implementation challenges posed by real-time data processing during cycling sessions. 

The salient strengths and contributions of this work encompass:

\begin{enumerate}
\item Versatility and Portability: 
Unlike present state of the art solutions that are fixed to a bicycle, the developed smart sensor node integrated into the cycling cleat is highly versatile and portable since a cycling cleat can be removed and mounted within two minutes. Unlike present solutions this design enables seamless use across various bikes. 
Furthermore, by reducing the hardware complexity and integrating off-the-shelf components, the total cost is limited to around 25\% of the price of the commercially available power meters that are based on torque-velocity measurement(e.g. Shimano from Stages Cycling). In addition, the manufacturing complexity, compared to a state of the art solution described in ~\cite{accuratePowerMeterDesign}, which relies on custom strain gauges, is significantly reduced. 
\item Real-time, low-power local Processing: 
By harnessing edge machine learning techniques in combination with a lightweight model (requiring only 146.8 kBytes of memory), the end-to-end signal inference latency is only 4.33 milliseconds on a microcontroller running at 84 MHz. A small 800 mAh 3.7 Volt lithium battery supports 25.8 hours of working time, resulting in a power efficiency of 1226 $\mu$Watt/MHz. The instantaneous feedback and long-time usage are thus ensured.
\item In-field Evaluation: 
In a final step, extensive multi-user in-field evaluation under real-life cycling conditions reinforces the validity and practicality of the proposed system. With a total of 830 minutes of usage, the hardware is rigorously tested and shows robust accuracy with a Mean Absolute Error (MAE) of 15.32 Watts (5.1\%), providing empirical evidence of its efficacy and reliability, which is only slightly bellow state of the art accuracy. 
\end{enumerate}

\section{Related Work}

Table \ref{relatedwork} briefly enumerates the previous endeavors on the topic of power estimation in cycling in recent years, presenting an overview of the utilized source signals and models. For instance, Mcainsh et al. T ~\cite{accuratePowerMeterDesign} built a crank arm embedded with strain gauge and IMU sensors and calculated the torque by the measured load combined with crank length. The power was derived from torque multiplied by rotational speed (angular velocity), the high precision measurement as well as intensive tuning of the sensors and custom design of the crank arm, which must be manufactured with high precision as its deformation is used for the force measurement. This allows a design that is capable to directly measure all 6 loads that may be applied to a crank arm. Thus achieves high accuracy of a mean accuracy of 2.5\%, as concluded in the work. Sate of the art sensors rely heavily on custom designed strain gauges incorporated in the crank arm. As described in ~\cite{accuratePowerMeterDesign} the gauge area is specially etched to achieve a perfect bond for the gauge and afterwards cured, ensuring a high precision gauge. In addition, increasing the number of strain gauges, allows further improvement of the device as the force is measured at a higher accuracy. This is described by Gardener et al. ~\cite{gardner204accuracy}. Burford et al.~\cite{PowerwithML} devised a cluster-based regression model employing input data such as heart-rate, cadence, and ridden kilometers to estimate average power during a bike ride, yielding a Mean Absolute Error (MAE) of 20 Watts. Oscar et al.~\cite{MLinCycling} proposed a Recurrent Neural Network (RNN) approach, leveraging time, speed, distance, altitude, power, and rider weight during cycling as input data to predict heart-rate response. Lemaitre et al.~\cite{PresentationPowerEstimation} proposed a deep neural network for power prediction, utilizing similar input data and claiming a Median Absolute Error of 25.1 Watts. In contrast to other research endeavours, our approach distinguishes itself by focusing on lightweight, energy-efficient, and low-latency Tiny Machine Learning (TinyML) models for edge processors, without the need for torque sensing and calculation. This shift allows us to optimize resource usage and process data more efficiently, thereby improving real-time power estimation during cycling. We also place a pronounced focus on IoT energy-efficient design, ensuring that our IoT device, CycloWatt, maintains extended battery life while delivering accurate real-time power estimations during cycling sessions. 
Besides the work of cycling power monitoring with AI solutions, similar research activities have been conducted for other fitness evaluations, such as wearable sensors on sports and workout profiling~\cite{bian2019passive, mekruksavanich2022multimodal, bian2022exploring}. These studies highlight the potential of deploying machine learning models to microcontrollers, maintaining accuracy in local inference compared to desktop or cloud computing. Our research builds upon these insights but directs its focus toward the specialized domain of power estimation in cycling.

\section{System Overview}

\begin{figure}[]
\begin{minipage}[t]{1.0\linewidth}
\centering
%\raggedright
\includegraphics[width=0.95\textwidth]{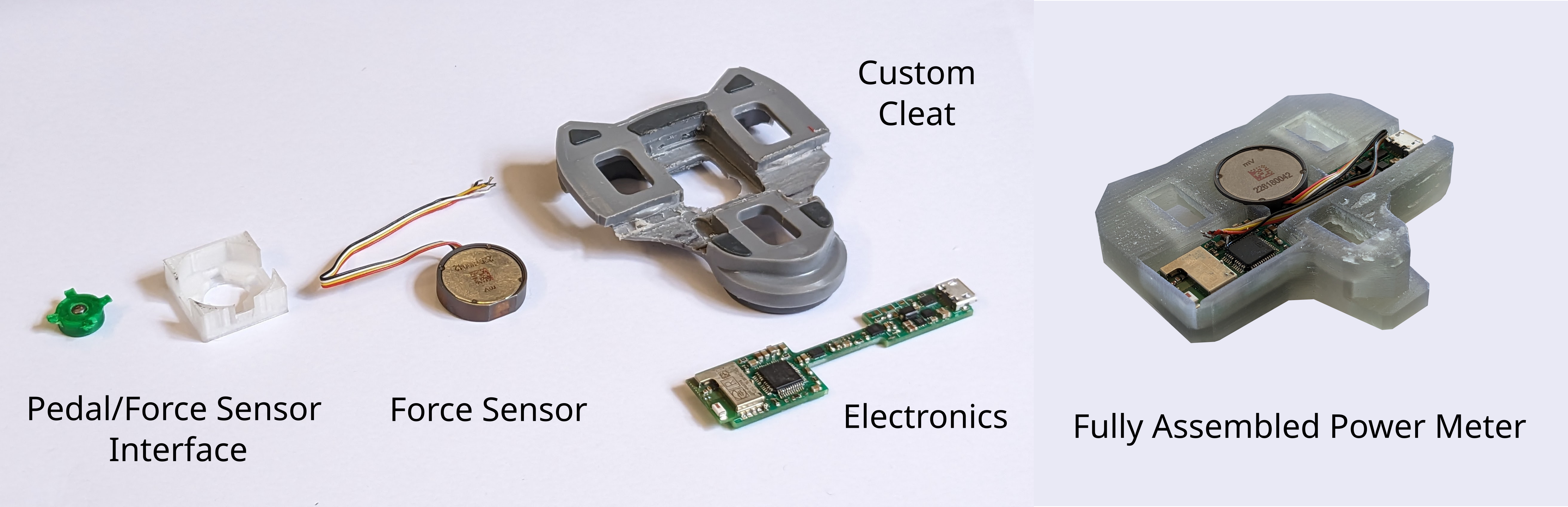}
\caption{Full disassembled hardware setup and final assembled device (Left to right: pedal to force sensor interface, force sensor, custom cleat, electronics and assembled power meter).}
\label{cleat}
\end{minipage}
\end{figure}

This section offers a hardware-software system overview of the proposed solution. It outlines two primary aspects: the design and integration of hardware components and the development and deployment of the tiny machine learning algorithm for cycling power estimation, utilizing only force and inertial signal inputs. 

\begin{figure}[]
\begin{minipage}[t]{1.0\linewidth}
\centering
%\raggedright
\includegraphics[width=0.85\textwidth]{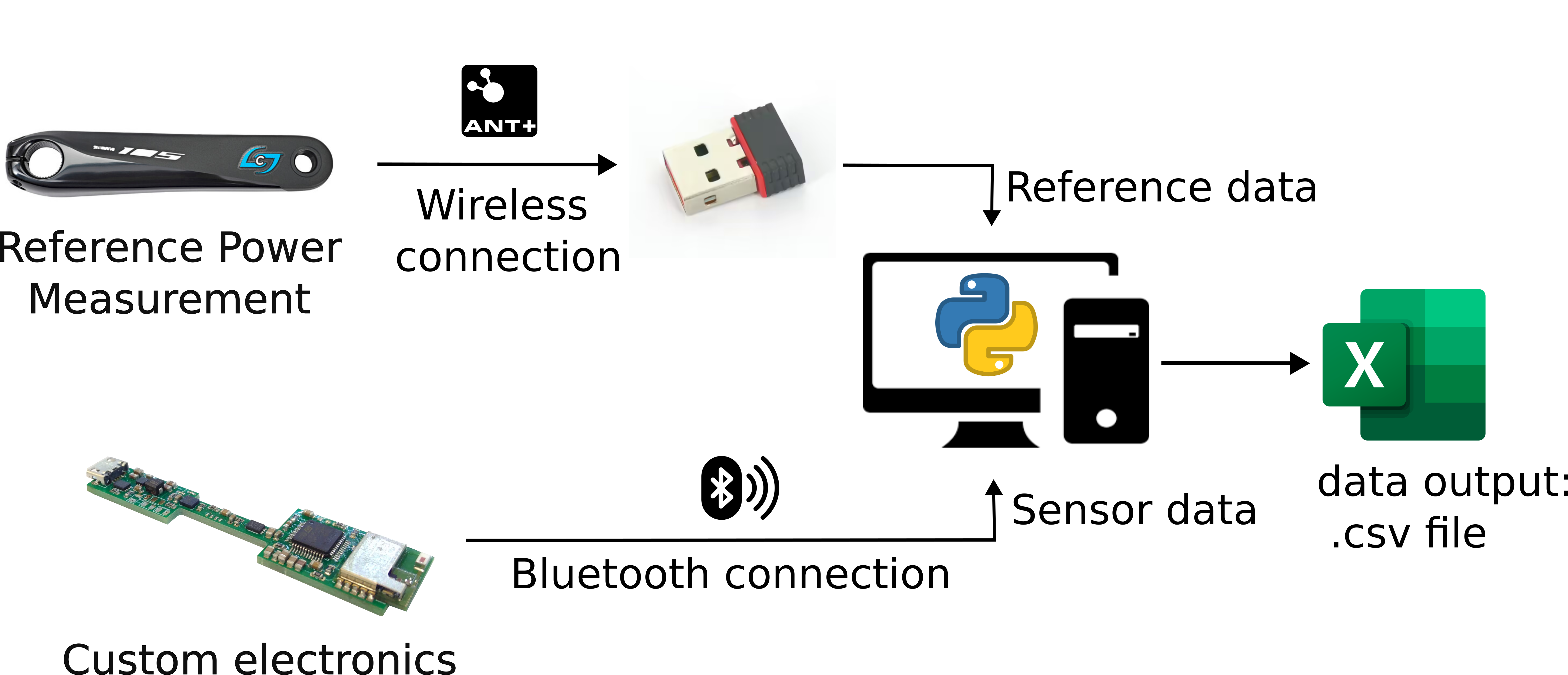}
\caption{Data collection setup with reference measurement. Illustration of the parallel collection and alignment of reference data and measured data.}
\label{ref_setup_overview}
\end{minipage}
\end{figure}

\subsection{Hardware}
The force sensor utilized in this system is a load cell based on resistive strain gauges arranged in a Wheatstone bridge configuration manufactured by TE Connectivity Measurement. This sensor offers a wide measurement range of up to 1000 Newton and boasts a compact design, with a diameter of 19 millimeters and a thickness of only 5 millimeters. Its output signal is an analog voltage in the millivolt range, necessitating the use of a high-impedance instrumentation amplifier for signal amplification and measurement. 
In conjunction with the force sensor, our sensor unit incorporates an additional Inertial Measurement Unit (IMU) from ST Microelectronics (LSM6DSOTR). The core electronics feature the STM32L4 microcontroller, which is part of the low-power series. This microcontroller is equipped with an ARM Cortex M4 processor and offers a maximum clock frequency of 120 MHz. Furthermore, it has 1024 kByte of Flash memory and 320 kByte of RAM, ensuring ample storage and processing capabilities.
To facilitate reliable and efficient communication, a low-power Bluetooth module (BLUENRG-M0L) is integrated into the custom PCB. This PCB is designed with a small total area of only 6.61 cm².

In order to improve the deployment convenience of the power meter, compared to commercially available options, all components are integrated into a cycling cleat. A cycling cleat can be exchanged by loosing three screws of the cycling shoe within minutes, compared to other options that are all integrated into the bike itself, requiring specialized tools, expertise, and time for deployment. An example is the R7000 power meter from Stages Cycling, which can only be mounted and unmounted with a wrench from Shimano. In the design phase, the paper presented by Mcainsh et al. T ~\cite{accuratePowerMeterDesign} served as a basis to decide on the needed measurements as well as the improvements that must be incorporated to the prototype compared to available solutions. 
Consequently, a custom-designed cycling cleat is meticulously developed to complete the prototype. Leveraging fast prototyping techniques, the cleat undergoes an iterative design to achieve a precise fit with the cycling pedal. This approach ensures an optimized interface between the cycling pedal and the force sensor. The ultimate cleat design is manufactured with a high-precision SLA printer from FormLabs. All the components of the prototype, showcasing the seamlessly integrated electronics, is visually depicted in Figure \ref{cleat}.

Nevertheless, the proposed hardware also has its merits concerning costs. The described setup amounts to less than seventy dollars, which is a significant reduction of 75\% compared to the price of the most affordable commercially available power meter (which is about three hundred dollars, offered by the Shimano series from Stages Cycling). Moreover, the installation and setup process for our solution is remarkably efficient, taking less than two minutes, which is approximately ten times faster than the current power meters available on the market. 

\subsection{Data Set and Data Collection Set-Up}
To train the edge machine learning model, a data set comprising force and inertial data is collected and synchronized with ground-truth power data derived from the Stages Cycling R7000. This power meter is installed at the left crank arm and claims an accuracy of up to 1.5\% \cite{AccuracyPowerMeterStagesCycling}, providing the resulting power via a wireless communication protocol. 
The experimental setup involves positioning the road bike on a stationary home trainer (Elite Novo Mag Force). This configuration allows users to adjust the wheel resistance as desired, enabling varied power levels while closely mimicking the outdoor cycling experience. 
To collect sensor data in real time and synchronize it with the reference measurement, a Python script is developed to acquire data from the custom electronics via Bluetooth, with a frequency of 58.3 Hz. Concurrently, a corresponding dongle facilitates the establishment of the ANT+ communication protocol, providing the reference measurement. 
The power meter furnishes ground-truth data at a rate of 4 Hz (corresponding to each pedal stroke). Figure \ref{ref_setup_overview} presents an overview of the setup. 

The primary objective of the data collection process is to ensure maximum variability, thereby cultivating a robust machine learning model. To achieve this, a detailed measurement protocol is devised, involving cycling at different power levels to obtain data points evenly distributed across the entire measurement range. This approach ensures the training of an unbiased model, free from any data set predispositions.
In addition to varying force levels, the cadence is also deliberately varied throughout the data acquisition sessions, as these two aspects play a significant role in different riding scenarios. It is imperative to encompass all possible combinations of high and low force levels and high and low cadence values. 
Each individual data acquisition session (5 minutes riding between 60 and 100 Watts, 100 and 140 Watts, 140 and 180 Watts, 180 and 220 Watts, 220 and 250 Watts, and above 250 Watts), adheres to the established protocol and spans approximately 30 minutes. \newline

\begin{figure}[]
\begin{minipage}[t]{1.0\linewidth}
\centering
%\raggedright
\includegraphics[width=0.7\textwidth]{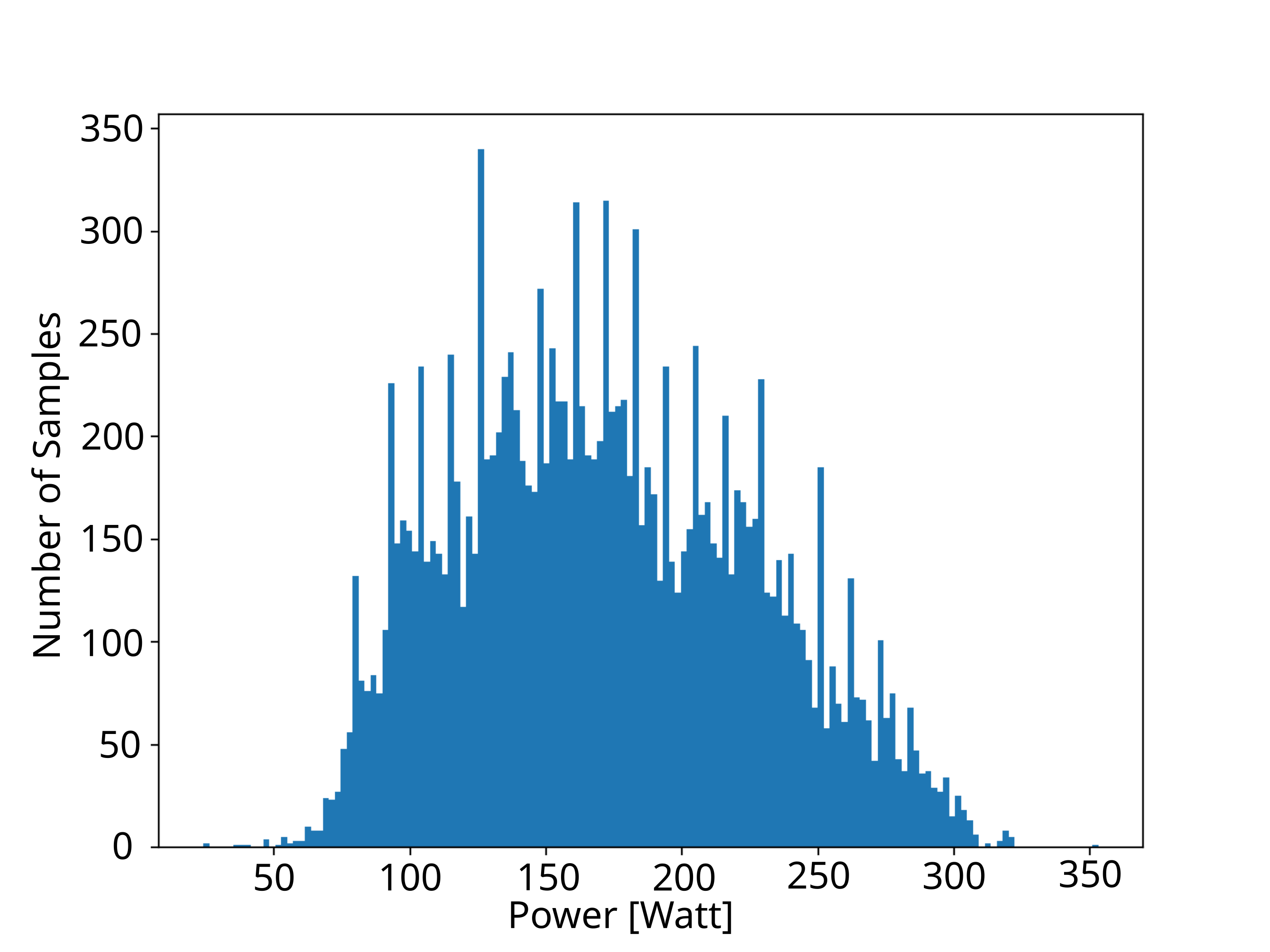}
\caption{Histogram of the collected training data, showing the distribution and balance of the training data.}
\label{histogram}
\end{minipage}
\end{figure}

To ensure the balance and diversity of the training data, a histogram (Fig. \ref{histogram}) is employed as a monitoring tool. This histogram showcases the distribution of collected training data across the entire power range, providing insights into potential deficits. This aids in identifying areas where data collection may be insufficient.
As the acquisition process progresses, this tool becomes instrumental in optimizing data collection. Any gaps or areas of limited data can be identified, prompting adjustments to the collection protocol. By targeting these specific ranges, the data collection process is fine-tuned to enhance the overall data balance.

The data gathering for this project involves repeating the described sessions over several weeks, resulting in approximately five hours of total riding and around 21,000 training samples spread over a measurement range of 300 Watts. The data collection and testing is performed by a male, amateur cyclist of age 27, between 60 and 70 kilograms and about 170 centimeters of height, engaging regularly in cycling activities.

\subsection{Data Processing}

The raw data from the force and inertial sensors must undergo pre-treatment for the machine learning model to estimate the cycling power precisely. This multi-stage process is summarized as follows: \newline 

\begin{enumerate}
\item Data Collection:
The measured force, linear acceleration along the X and Z-axis, and the angular acceleration around the Y-axis undergo low-pass filtering and are subsequently stored with a sampling frequency of 58.3 Hz to a ring buffer. The mentioned axes are measured because those are the degrees of freedom a cycling pedal has. Thus, we may expect the most significant movements along these axes, providing a unique movement pattern depending on the power level of a pedal stroke. 
\item Pedal Stroke Detection:
The data in the ring buffer is perpetually monitored for maximum points in the force channel, as these points signify the initiation and conclusion of pedal strokes. This enables the extraction and segmentation of individual pedal stroke candidates.
\item Pedal Stroke Separation:
The previously detected pedal stroke candidates are subjected to validation to ensure they adhere to general properties, such as force shape and amplitude, indicative of a genuine pedal stroke. This process allows the removal of false candidates. Subsequently, the data is divided into individual pedal strokes and adjusted to a predefined length of 32 samples, ensuring uniformity.
\item Feature Extraction:
Additional features are extracted from each pedal stroke sample, encompassing cadence, absolute force amplitude, and force offset. Feature extraction is imperative to retain essential information, as normalization and data resizing otherwise lead to loss of critical details.
\item Normalization:
In the final step, the data is normalized to rescale the collected data within a range of zero and one. This normalization process is essential to enhance accuracy, ensuring all samples maintain a consistent scale. 
\end{enumerate}

\subsection{Proposed TinyML Model}
For the design of the employed machine learning model the paper from Lemaitre et al.~\cite{PresentationPowerEstimation} is used as an initial guiding for the composition of a lightweight model structure tailored to meet the strict memory constraints imposed by the microcontroller. The network consists of three fully connected hidden layers (256, 128, 32 neurons, separately), and its specific architecture is determined through an iterative process. 
The resulting neural network comprises 70,337 trainable parameters and a total of 546 neurons. Utilizing Relu activation functions, the model is trained for 256 learning epochs, incorporating early stopping with a patience of 5 and a batch size of 128 samples. The learning rate is set to 0.01 with a time-dependent decaying rate, while the loss function used is the Mean Squared Error. Overall, the model requires 270.5 kBytes of memory with full precision accuracy (32-bit floating point). To optimize memory utilization, the model undergoes quantization to 8-bit precision, reducing the memory requirement by nearly a factor of four to a mere 71.9 kByte. The entire model development process is carried out in a Jupyter Notebook, leveraging the Keras library as a wrapper. 

\section{ Experimental Evaluation}
The final prototype undergoes rigorous testing to assess its performance and reliability involving several stages: initial evaluation examines the machine learning model on a desktop computer using recorded data, followed by indoor tests on a roller trainer, and culminating in in-the-wild tests under real-world conditions. This incremental approach allows for a conclusive understanding of the system's behavior and enables the monitoring of accuracy under various influences, fostering a step-by-step understanding of the system and making it possible to pinpoint potential issues. The subsequent section outlines in detail the obtained test results.

\subsection{Prototype Testing}
\begin{figure}[!t]
\centering
\begin{minipage}[t]{1.0\linewidth}
%\raggedright
\centering
\includegraphics[width=1\textwidth]{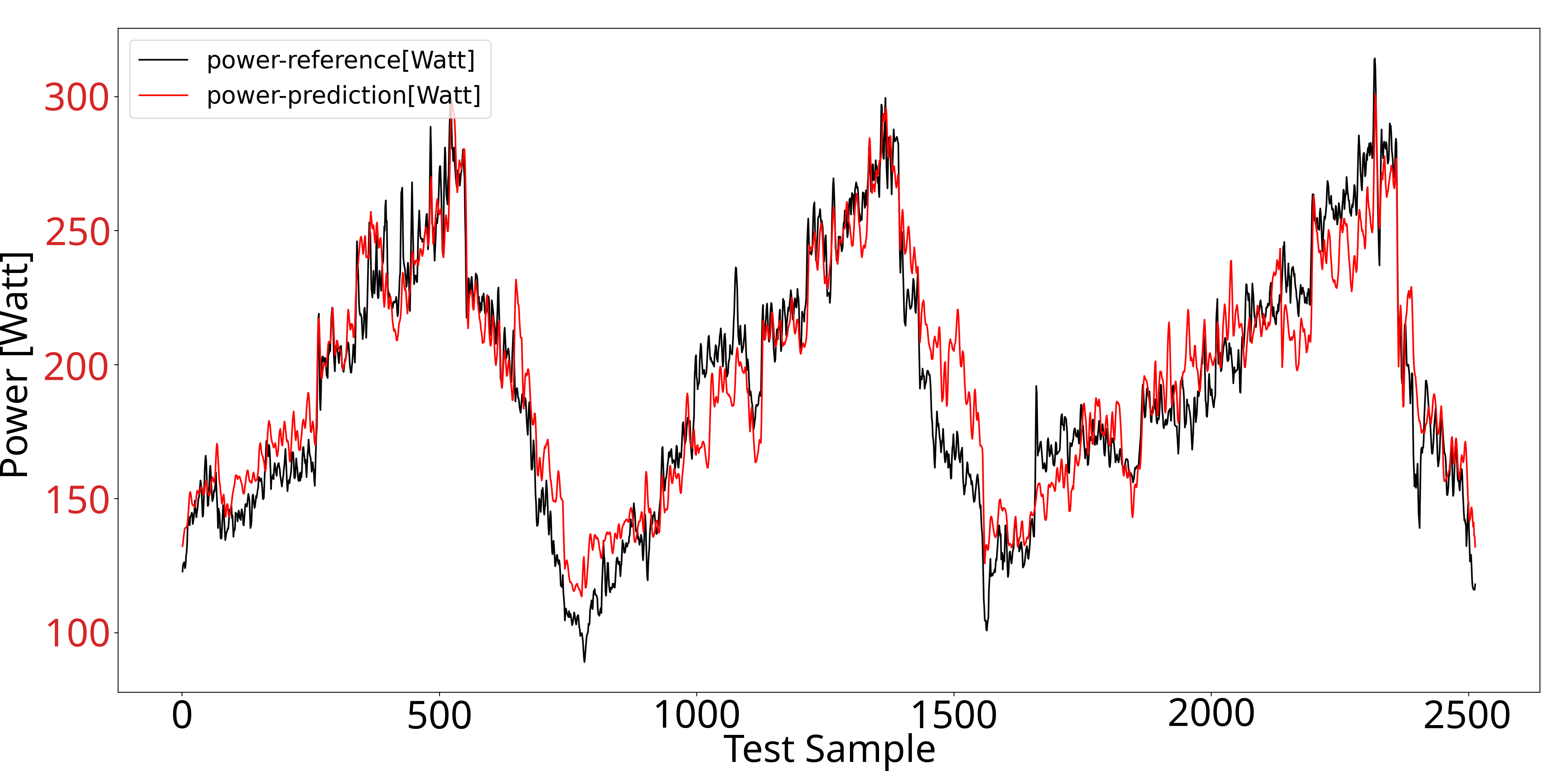}
\end{minipage}
\caption{Results of the three indoor test rides, using the unquantized model.This compares the reference power measurement (black) with the estimated power (red).}
\label{unquantized_results}
\end{figure}

The prototype testing started off with a desktop evaluation of the proposed machine learning model. This is done using pre-recorded testing data consisting of a total of 4,250 samples. Resulting in an error of 13.28 Watts Mean Absolute Error (MAE), giving initial insights about the machine learning model. In the next testing phase, the unquantized machine learning model is deployed to a microcontroller and tested on a roller trainer for a total of 30 minutes. This test included the first use of the prototype hardware and was spread over three individual testing rides, each of approximately ten minutes. Yielding a total of 2,500 power samples estimated by the model. This test provides profound initial insights into the full system's potential and behavior. After all the error of this test is measured to be 14.309 Watts (MAE). A graphical overview of the test result is provided in Figure \ref{unquantized_results}.  

\begin{figure}[!t]
\centering
\begin{minipage}[t]{1.0\linewidth}
%\raggedright
\centering
\includegraphics[width=1\textwidth]{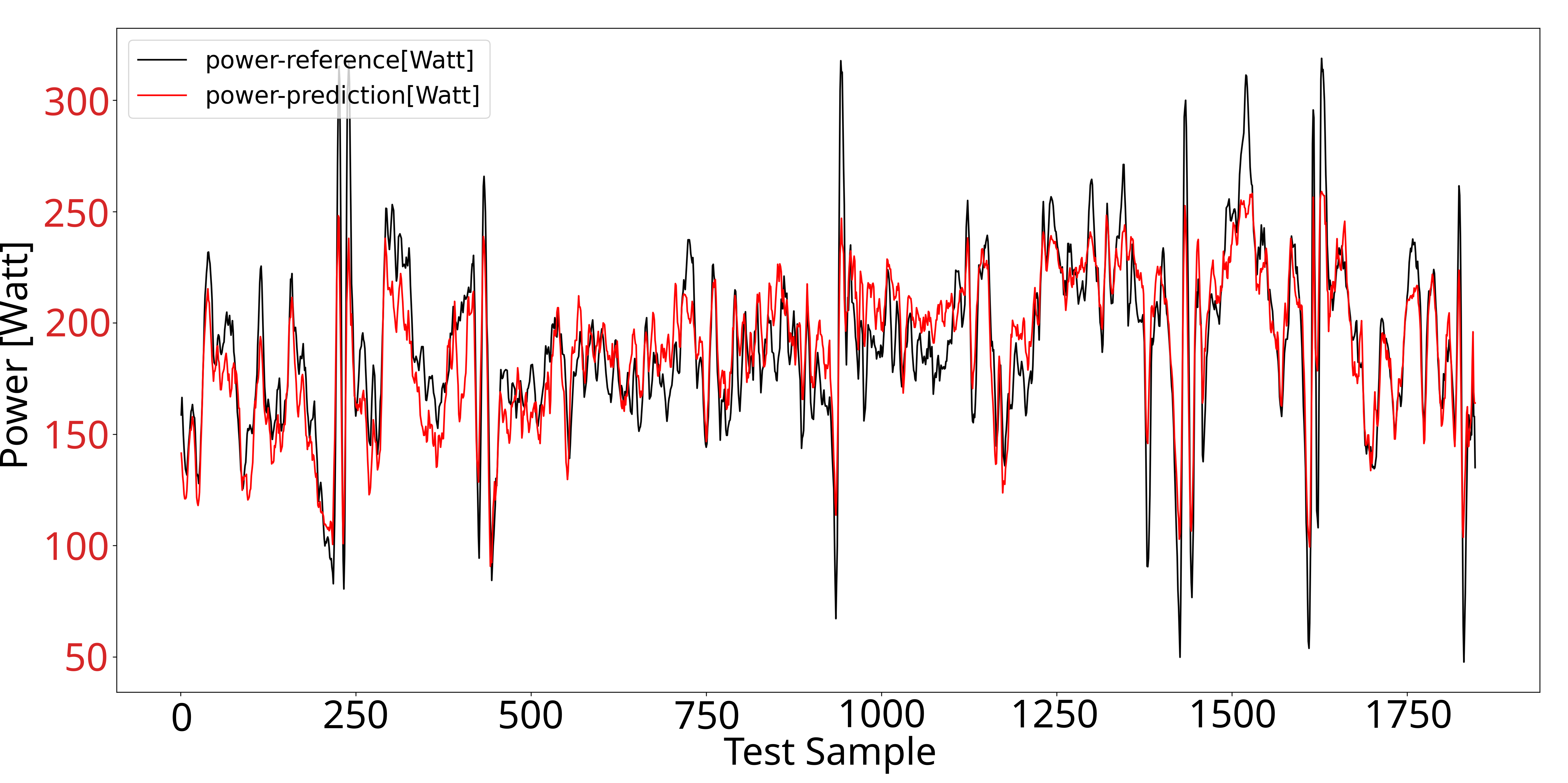}
\end{minipage}
\caption{Results of the 11.3 kilometer outdoor test ride, with reference power (black) and estimated power (red).}
\label{outdoor_test}
\end{figure}

In order to optimize the model for use on a microcontroller, the model is quantized to 8-bit precision, first tested on a desktop computer and thereafter evaluated on a roller trainer. The power estimation is performed for each pedal stroke in real-time during a 30-minute indoor bike ride, generating a total of 1,570 samples. The Mean Absolute Error for this test is determined to be 12.289 Watts and the average power is estimated with a difference of only 1.735 Watts for the entire ride. Subsequently, the same machine learning model is employed to estimate power during an outdoor ride covering a distance of 11.2 kilometers over a duration of 25 minutes, yielding 1,850 samples. The results are depicted graphically in Figure \ref{outdoor_test}. Remarkably, this real-life, in-the-wild test, which exposes the prototype to various external influences, produces a superior outcome with a MAE of only 15.321 Watts. The average power estimation during the entire ride exhibits a difference of merely 0.889 Watts compared to the reference.

To validate the model's generalizability, eight different riders of varying ages, genders, and fitness levels partake in testing the prototype for 12 minutes on the roller trainer across various power levels encompassing the entire power range. These tests account for a total of 96 minutes and 8,150 test samples, resulting in an average error of 15.947 Watts (MAE). Table \ref{result} offers a comprehensive overview of the conducted tests with the prototype. In summary, the extensive and meticulous testing, spanning multiple extended sessions with a combined duration of 830 minutes, underscores the remarkable efficacy of fusing force and inertial signals for cycling power estimation. Moreover, the prototype's consistent and intensive usage attests to its reliability and durability.

\begin{table}[t]
\centering
\footnotesize
\caption{Prototype testing with 8-bit quantized model}
\label{result}
\begin{tabular}{p{1.5cm}|p{1.6cm}|p{1.6cm}|p{1.6cm}} 
 \hline
 \textbf{Metric} & \textbf{Roller trainer test (quantized)} &  \textbf{Outdoor test} & \textbf{Generalization test}  \\ 
 \hline
 Test samples & 1,570  & 1,850  & 8,150   \\
 \hline
 Riding time & 20 minutes  & 25 minutes  & 96 minutes   \\
 \hline
 Mean Absolute Error & 12.289 Watts & 15.321 Watts & 15.947 Watts \\
\hline
\end{tabular}
\end{table}

\subsection{Power Consumption and Latency}
The hardware is also subjected to tests assessing its latency and power consumption. These insights may also provide valuable information about the battery life, validating the estimated operational duration of the prototype. The InfiniiVision DSO-X 4024A oscilloscope from Agilent Technologies is utilized to measure the power consumption in both active mode and low power mode. During active mode the device only requires a total of 0.103 Watts, which is the basis for a long battery operation up to 25 hours (800 mAh battery). The detailed results of this evaluation are summarized in Table \ref{result_hardware}.

\begin{table}[]
\centering
\footnotesize
\caption{Hardware characterization (Power Consumption)}
\label{result_hardware}
\begin{tabular}{ p{4.2cm}|p{3.0cm}} 
 \hline
 \textbf{Measurement} & \\ 
 \hline
 Average power  & 0.103 Watt \\
 \hline
 Peak power & 0.148 Watt \\ 
 \hline
 Energy efficiency & 0.177 Joule/estimation\\
 \hline
 Power efficiency &  1226 $\mu$Watt/MHz \\
  \hline
 Average power (low power mode)&  0.0507 Watt \\
 \hline
\end{tabular}
\end{table}

Furthermore, an analysis of the latency for different tasks is conducted. This involves measuring the number of clock cycles required to complete individual code sections, which in turn determines the latency. Table \ref{result_hardware_latency} provides an overview of the hardware's latency performance.
Throughout all the performed tests, the clock frequency is set to 84 MHz, which is a trade-off between performance and power consumption. 

\begin{table}[]
\centering
\footnotesize
\caption{Hardware characterization (Latency)}
\label{result_hardware_latency}
\begin{tabular}{ p{3.7cm}|p{3.5cm}} 
 \hline
 \textbf{Latency @ 84 MHz} & \\ 
 \hline
 Pre-processing  & 0.824 milliseconds \\
 \hline
 Post-processing & 3.1785 microseconds \\ 
 \hline
 Sending data & 0.169 milliseconds\\
 \hline
 Inference & 3.333 milliseconds \\
 \hline
  \hline
 Total latency per estimation &  4.33 milliseconds \\
 \hline
\end{tabular}
\end{table}

\section{Conclusion}
This paper introduced CycloWatt, an affordable, TinyML-enhanced IoT device revolutionizing cycling power metrics.
The extensive usage of the CycloWatt prototype during testing ensured a profound evaluation of every aspect of its functionality. Through various tests, the prototype’s performance is assessed on a roller trainer, with different riders, and in outdoor in-the-wild scenarios. The achieved results demonstrate strong competitiveness when compared to previous works in the field of machine learning-based power estimation in cycling. 
Future endeavors should primarily focus on enhancing the mechanical setup and further developing the custom cleats, while the electrical setup requires only minor adjustments. Additionally, tests have revealed the potential of a machine-learning model trained on an improved data set. Consequentially, future iterations of the prototype should prioritize data collection from various riders to further enhance the model’s performance. Overall, this paper lays a solid foundation by showcasing the prototype’s capabilities, emphasizing the significance of leveraging machine learning for accurate and cost-efficient power estimation in cycling. 

\bibliographystyle{IEEEtran}
\bibliography{sample.bib}

\end{document}